\documentclass[12pt]{iopart}
\usepackage{graphicx}

\begin{document}

\title{Threshold meson production and cosmic ray transport}

\author{John W. Norbury}

\address{Department of Physics and Astronomy, University of Southern Mississippi, Hattiesburg, Mississippi, 39402, USA}
\ead{john.norbury@usm.edu}

\author{Lawrence W. Townsend}

\address{Department of Nuclear Engineering,
University of Tennessee, Knoxville, TN 37996}
\ead{ltownsen@tennessee.edu}

\author{Ryan B. Norman}

\address{Department of Physics,
Worcester Polytechnic Institute, Worcester, MA 01609}
\ead{rbnorman@wpi.edu}

\date{\today}

\begin{abstract}
An interesting accident of nature is that the peak of the cosmic ray spectrum, for both protons and heavier nuclei, occurs  near the pion production threshold. The Boltzmann transport equation contains a term which is the cosmic ray flux multiplied by the cross section. Therefore when considering pion and kaon production from proton-proton reactions, small cross sections at low energy can be  as important as larger cross sections at higher energy.   This is also true for subthreshold kaon production in nuclear collisions, but not for subthreshold pion production.

\end{abstract}


\section{Introduction}

 The existence of the pion was predicted by Yukawa  \cite{yukawa}   in 1935 and discovered in the comsic ray spectrum by Powell  \cite{powell} in 1947. Seventy years later one would have thought that of all reactions,  pion production in proton proton collisions was thoroughly understood.
There has been considerable interest recently in meson production from nucleon nucleon collisions near the meson production threshold \cite{hanhart, dos santos, machner, moskal}. Pion production has been studied in the most detail. The reason these studies raised such interest was because the first calculations of total cross sections disagreed with experiment by a factor of five \cite{machner} and the differential cross sections for spin observables had  the wrong shape, being convex instead of concave \cite{meyer}.

Particle reaction transport  codes, such as GEANT \cite{geant} and FLUKA \cite{fluka} are widely used in the design of accelerator experiments and the simulation of particle detectors.
While typical cross sections away from threshold reach tens of millibarn, the cross sections near threshold are about a thousand times smaller, typically in the microbarn region. The fact that theory and experiment disagreed so much in the threshold region was therefore of no real concern for particle simulation codes because the cross sections in the threshold region are so  small.

Transport codes are also widely used in studies of cosmic rays.
If one knows the  cosmic ray spectrum incident on top of the Earth's atmosphere, then one can deduce the cosmic ray spectrum observed on the ground by transporting through the atmosphere. Similarly if  one knows the  cosmic ray spectrum incident on a spacecraft wall then one can deduce the radiation environment inside a spacecraft \cite{wilson}. In cosmic ray research one turns this around to deduce the incident spectrum from knowledge of the ground based spectrum.

Transport codes are  concerned with solving the Boltzmann transport equation. In the continuous slowing down and  straight-ahead approximations, the one dimensional Boltzmann equation can be expressed \cite{wilson}  as 
\begin{eqnarray}
\left [ \frac{\partial}{\partial x} - \frac{\partial}{\partial E} S_j(E) + \Sigma_j(E) \right ] 
\phi_j(x,E)  
= \int dE^\prime \sum_k \Sigma_{jk}(E, E^\prime) \phi_k(x, E^\prime)
\end{eqnarray}
where S(E) is the stopping power, $\Sigma$ is  the macroscopic  cross section  and $\phi$ is the flux of particles at a given depth $x$ and energy $E$.
The macroscopic cross section is related to the microscopic cross section $\sigma$ by $\Sigma = \rho \sigma$, where $\rho$ is the target number density. Note that the flux $\phi$ is multiplied by the cross section $\sigma$. 
In a particle accelerator, the incident flux is normally a beam (constant flux) of particles with a well defined energy. In cosmic ray physics the incident flux is represented by the incident cosmic ray spectrum, which is discussed in the review by Simpson \cite{simpson}. In Figure 5a of that review one can find the incident spectrum for protons and other nuclei. The peak of the spectrum occurs at an energy of about 300 MeV. The peak is caused by solar modulation of the interplanetary magnetic field and is known to move around somewhat depending on solar activity. Note  that the peak of the spectrum occurs quite near the pion production threshold of 290 MeV. This is an interesting accident of nature. Above  this energy the spectrum starts falling. 

Now  the Boltzmann equation contains flux multiplied by cross section.  Therefore  just because a cross section is small does not mean that it is unimportant. A small cross section might get multiplied by a large flux and a large cross section might get multiplied by a small flux, leading to approximately equal contributions from a transport point of view. This is possible  when the incident flux is the cosmic ray spectrum at the top of the Earth's atmosphere.

\section {Pion and kaon production in proton-proton collisions}

Consider the exlusive reaction for production of neutral pions,
\begin{eqnarray}
p + p \rightarrow p + p +\pi^0
\end{eqnarray}
Sample values of cosmic ray proton flux  \cite{simpson} and  microscopic cross section are given in Table 1, where the values are also multiplied together (we call this  {\em importance}), as in the Boltzmann equation. It can be seen  that the small cross sections near threshold can be of similar importance to larger cross sections away from threshold, the reason being due to the fact that the cosmic ray spectrum has a peak near threshold and falls steadily at higher energies.  For example, the cross section at 375 MeV is 40 $\mu$b, while the cross section at  7  GeV is 1.7 mb, yet they give approximately equal contribution to the importance factor.  
Of course these do not compare to the importance of the cross sections at say 700 MeV and 1 GeV. However in a good transport code one will always include the cross section at a variety of  energies and one would certainly go out to 7 GeV. The point is that if one does include such higher energy cross sections, then one may also need to  include the cross sections near threshold as well because they are of comparable importance.\\

\noindent {\bf Table  1.}  Neutral pion production from proton-proton collisions. (Importance $\equiv$ Flux $\times\; \sigma$.) The numbers in square brackets after the cross sections indicate the Reference number from where the cross section was taken.  The peak in  Flux  occurs near the pion  threshold.

\noindent
\hrulefill
\begin{tabbing}
xxxxxxxxxxxxxxxxxxx\=xxxxxxxxxxxxxxxxxx\=xxxxxxxxxx\=xxxxxxxxxxxxx\=xxxxxxxxxxx\kill
T\> Flux  \cite{simpson} \>$\sigma $ \>Ref. \>Importance \\
(MeV)\>($\frac{1}{ \rm m^2 \; sr \; s\;  MeV}$) \>($\mu {\rm b}$)\>
\>($\frac{\mu {\rm b}}{ \rm m^2 \; sr \; s\;  MeV}$)\\

\end{tabbing}
\hrulefill

\begin{tabbing}
xxxxxxxxxxxxxxxxxxx\=xxxxxxxxxxxxxxxxxx\=xxxxxxxxxx\=xxxxxxxxxxxxx\=xxxxxxxxxxx\kill
290 $\leftarrow$  {\em  threshold }\\
325          \>2.0            \>7.7     \> \cite{meyer}           \>15.4           \\
350          \>2.0            \>17      \> \cite{meyer}        \>34     \\
375          \> 1.8           \>40      \> \cite{meyer}          \>72           \\
400          \>1.8             \>86    \> \cite{meyer}         \> 155        \\
700          \>1.3            \>2000    \>  \cite{teis}       \>2600           \\
 1000      \>0.9            \>4000    \>  \cite{teis}          \>3600           \\
3000          \>0.24          \>3000    \>\cite{teis}        \>720           \\
7000         \> 0.06          \>1700     \>  \cite{teis}     \>102          \\
11500          \> 0.01           \>1100     \> \cite{teis}      \> 11          
  \end{tabbing}
\hrulefill  \\

Similar statements can be made concerning kaon production as shown in Table 2, where the lowest threshold  reaction for producing kaons,
\begin{eqnarray}
p + p \rightarrow K^+ + \Lambda + p
\end{eqnarray}
is considered.
It can be seen that the importance of near-threshold reactions is  comparable to reactions at all energies.
Therefore it may be  important to consider near-threshold production of kaons.\\

\noindent    {\bf Table  2.}  Kaon production from  $pp\rightarrow K^+ \Lambda p$.    

\noindent
\hrulefill
\begin{tabbing}
xxxxxxxxxxxxxxxxxxx\=xxxxxxxxxxxxxxxxxx\=xxxxxxxxxx\=xxxxxxxxxxxxx\=xxxxxxxxxxx\kill
T\> Flux  \cite{simpson} \>$\sigma $ \>Ref. \>Importance \\
(GeV)\>($\frac{1}{ \rm m^2 \; sr \; s\;  MeV}$) \>($\mu {\rm b}$)\>
\>($\frac{\mu {\rm b}}{ \rm m^2 \; sr \; s\;  MeV}$)

\end{tabbing}
\hrulefill

\begin{tabbing}
xxxxxxxxxxxxxxxxxxx\=xxxxxxxxxxxxxxxxxx\=xxxxxxxxxx\=xxxxxxxxxxxxx\=xxxxxxxxxxx\kill
1.58 $\leftarrow$  {\em  threshold }\\
1.60        \>0.6             \>0.16            \> \cite{balewski}           \>0.1            \\
1.82          \>0.5            \>7.4              \> \cite{abdel}          \>3.7           \\
 1.90         \>0.4           \> 8.6              \>  \cite{abdel}        \>3.4           \\
2.06          \>0.3            \>16.5            \> \cite{abdel}            \>5.0           \\
  2.85         \>0.2             \> 50             \> \cite{fuchs}         \>10            \\
  4.83         \>0.1             \> 50             \>\cite{fuchs}          \>5            \\
  7.29          \>0.06             \> 50          \>\cite{fuchs}             \>3            \\
30         \>.002            \> 16.6              \> \cite{cleland}        \> 0.03          \\
  50        \>0.0005            \> 16.7             \> \cite{cleland}         \> 0.01

  \end{tabbing}
\hrulefill  \\

\section{Sub-threshold pion and kaon production in nucleus-nucleus collisions}

Based on the above considerations, one might ask about the subthreshold pion production observed in nucleus-nucleus collisions some decades ago  \cite{norbury}. Here the word subthreshold refers to pions produced in nuclear collisions where the nucleus kinetic energy is below the nucleon-nucleon threshold. The typical way to express energies in nucleus-nucleus collisions is using A MeV.  An energy of   290 A MeV  would represent the nucleon-nucleon threshold. Simpson \cite{simpson} also provides the cosmic ray spectrum for nuclei, such as He, C, Fe. 
Again the spectrum peak for these nuclei is right near the pion threshold. If the peak were  below the pion threshold then sub-threshold pion production in nuclear collisions would be important for the same reasons as given above. However because the nuclear peaks remain near the pion threshold then subthreshold pion production is not important in the sense described previously. This can be seen from Table 3 where total cross sections for neutral pion production in Carbon-Carbon collisions below the pion threshold are presented.  Due to a lack of total cross section data, the experimental cross sections without Reference numbers next to them were    taken from Table 1 and multiplied by 144, which is the product of the mass numbers of the projectile and target nuclei.   Therefore they are approximate only.
Again these cross sections are multiplied by the Carbon cosmic ray flux and it can be seen that  sub-threshold cross sections are not important compared to cross sections at higher energies. Therefore it is probably safe to neglect sub-threhsold pion production in transport codes.\\

\noindent    {\bf Table  3.}    $\pi^0$  production from C-C  collisions. 
  Note that the  peak in  Flux  occurs near the pion  threshold.

\noindent
\hrulefill
\begin{tabbing}
xxxxxxxxxxxxxxxxxxx\=xxxxxxxxxxxxxxxxxx\=xxxxxxxxxx\=xxxxxxxxxxxxx\=xxxxxxxxxxx\kill
T\> Flux  \cite{simpson} \>$\sigma $ \>Ref. \>Importance \\
(AMeV)\>($\frac{1}{ \rm m^2 \; sr \; s\;  AMeV}$) \>($ {\rm mb}$)\>
\>($\frac{ {\rm mb}}{ \rm m^2 \; sr \; s\;  AMeV}$)

\end{tabbing}
\hrulefill

\begin{tabbing}
xxxxxxxxxxxxxxxxxxx\=xxxxxxxxxxxxxxxxxx\=xxxxxxxxxx\=xxxxxxxxxxxxx\=xxxxxxxxxxx\kill
60          \>$5 \times 10^{-3}$            \>$1.7  \times 10^{-3}$            \>\cite{noll, braun-munzinger}          \>8.5  $\times 10^{-6}$         \\
74          \>$6 \times 10^{-3}$            \>$8.5   \times 10^{-3}$  \>\cite{noll, braun-munzinger}         \>  5.1  $\times 10^{-5}$        \\
84         \>$7 \times 10^{-3}$           \>$18.9   \times 10^{-3}$              \>\cite{noll, braun-munzinger}          \>1.3  $\times 10^{-4}$           \\
290 $\leftarrow$  {\em  threshold }\\
325          \>$7 \times 10^{-3}$            \>1     \>           \>0.007           \\
350          \>$7 \times 10^{-3}$             \>2      \>       \>0.01           \\
375          \>$6 \times 10^{-3}$             \>6      \>         \>0.04           \\
400          \> $6 \times 10^{-3}$            \> 12   \>        \>0.07          \\
700          \> $4 \times 10^{-3}$            \>288    \>        \>1.2           \\
 1000      \>$3 \times 10^{-3}$             \> 576   \>            \>1.7           \\
3000          \> $5 \times 10^{-4}$           \> 432   \>        \>0.2           \\
7000         \>$1 \times 10^{-4}$             \> 245    \>       \>0.02           \\
11500          \> $2 \times 10^{-5}$            \>158     \>      \>0.003           
  \end{tabbing}
\hrulefill \\

It is interesting to look at subthreshold kaon production because  the kaon threshold is well above the pion  threshold. Data are presented in Table 4.  Again  due to a lack of total cross section data, the 
  experimental cross sections without Reference numbers next to them were  taken from Table 2 and multiplied by 3364, which is the product of the mass numbers of the projectile and target nuclei. It can be seen that subthreshold cross sections are as important as cross sections near 10 GeV.  Therefore one can make the interesting conclusion that  subthreshold kaon production may need to be included  in transport codes. 
  A better knowledge of the nucleus-nucleus total cross sections is necessary in order to make more definitive statements.

\newpage

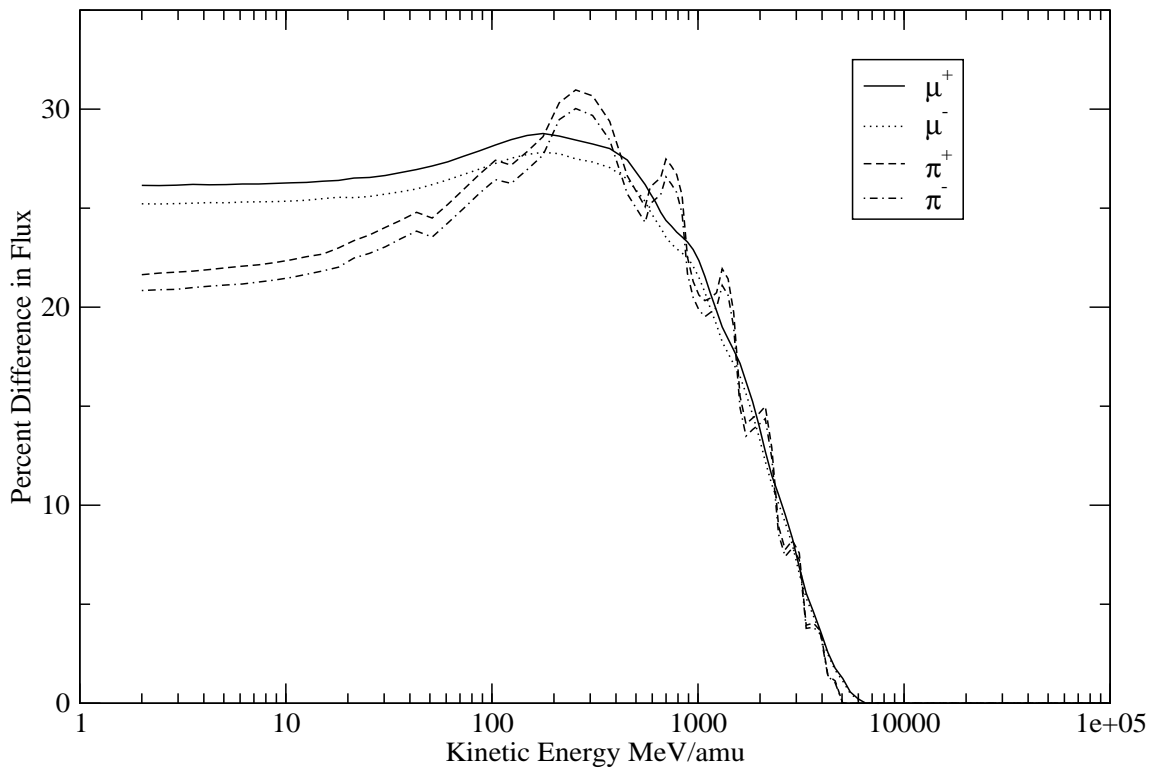
\noindent {\bf Figure 1.} The percent flux difference of charged muons and pions in the Mars atmosphere \cite{angelis1} at a depth of 38.9 $\frac{g}{cm^2}$ from the 1977 GCR spectrum \cite{badwar1}  

  \section {An example transport calculation}
  We have added this section on the advice of a referee, who wished us to do a sample transport calculation by artificially multiplying the cross section near threshold by a factor of two to see what the effects might be. The transport calculation was done with MESTRN \cite{mestrn}, an extension of the NASA transport code HZETRN \cite{hzetrn} that includes the production of charged pions and muons.  One improvement that was made to the published version of MESTRN is the pion production spectral distributions from proton-proton collisions are now calculated by numerically integrating the Lorentz invariant differential cross sections of Badwar {\it et al.} \cite{badwar1}.  This was done to achieve better numerical convergence for low density materials and to allow for transport through variable density materials.  

The Martian atmospheric model of De Angelis {\it et al.} \cite{angelis1} and the primary galactic cosmic ray (GCR) model of Badwar and O'Neill \cite{badwar2} were used as inputs for the transport calculation.  Two runs of the transport code were performed.  For one run, the cross section for the production of pions was artificially multiplied by a factor of two from threshold to a kinetic energy of 2 GeV.  

\begin{figure}[tb]
\resizebox{6in}{4in}{
\includegraphics{fig1.eps}}
\end{figure}

Figure 1 shows the percent flux difference of charged pions and muons after transport through
38.9 $\frac{g}{cm^2}$ of Martian atmosphere corresponding to an approximate altitude of -10 km.  A negative altitude indicates a valley or impact basin.  The percent flux difference is the flux of the MESTRN run with the cross section for pion production enhanced by a factor of two minus the nonenhanced flux.  This value is then divided by the nonenhanced flux to give the percent variation due to the enhancement of the pion production cross section near threshold.

The enhanced pion production cross sections led to an enhanced flux of charged pions and muons by approximately 20-30\% in the kinetic energy region up to 1 GeV. The enhancement fades quickly after 1 GeV.  We might expect this effect to be larger, but the produced pions decay quickly into muons which subsequently decay also.  Muon decay in MESTRN is treated simply as a loss term.  Subsequently, some of the muons that are produced from the increased pion production near threshold are not accounted for in these results.\\

\noindent    {\bf Table  4.}     $K^+$ production from Ni-Ni  collisions. 
 The approximate total experimental   cross sections are from Reference \cite{barth}. 
 To obtain the flux for Ni, the flux of Fe was used but multiplied by 0.05, which is the average ratio of the abundance of Ni to Fe over all energy intervals \cite{simpson}.

\noindent
\hrulefill
\begin{tabbing}
xxxxxxxxxxxxxxxxxxx\=xxxxxxxxxxxxxxxxxx\=xxxxxxxxxx\=xxxxxxxxxxxxx\=xxxxxxxxxxx\kill
T\> Flux   \>$\sigma $ \>Ref. \>Importance \\
(AGeV)\>($\frac{1}{ \rm m^2 \; sr \; s\;  AMeV}$) \>($ {\rm mb}$)\>
\>($\frac{ {\rm mb}}{ \rm m^2 \; sr \; s\;  AMeV}$)

\end{tabbing}
\hrulefill

\begin{tabbing}
xxxxxxxxxxxxxxxxxxx\=xxxxxxxxxxxxxxxxxx\=xxxxxxxxxx\=xxxxxxxxxxxxx\=xxxxxxxxxxx\kill
0.8          \>     1.1  $  \times 10^{- 5 }$        \>.9              \> \cite{barth}         \>  1.0 $  \times 10^{- 5 }$     \\
1.0          \>  1 $  \times 10^{- 5 }$            \>2.7             \> \cite{barth}         \>  2.7 $  \times 10^{- 5 }$     \\
1.58 $\leftarrow$  {\em  threshold }\\
1.60        \>   1 $  \times 10^{-5  }$          \>.5            \>            \>  5 $  \times 10^{- 6 }$   \\
1.8          \>   5 $  \times 10^{- 6 }$           \>57              \>\cite{barth}          \> 2.9$  \times 10^{- 4 }$          \\
1.82          \>  5$  \times 10^{- 6 }$            \> 25             \>          \>  1.3$  \times 10^{- 4 }$     \\
 1.90         \>  4.5$  \times 10^{- 6 }$          \> 29              \>          \> 1.3$  \times 10^{- 4 }$       \\
2.06          \>   3.5$  \times 10^{- 6 }$        \>56            \>             \>  2.0$  \times 10^{- 4 }$       \\
  2.85         \> 2.5$  \times 10^{- 6 }$            \>168              \>         \> 4.2$  \times 10^{- 4 }$        \\
  4.83         \>  1$  \times 10^{- 6 }$             \>168              \>        \> 1.7$  \times 10^{- 4 }$        \\
  7.29          \>  4$  \times 10^{- 7 }$            \>168           \>             \>  6.7$  \times 10^{- 5 }$        \\
30         \>   2.5$  \times 10^{- 8 }$        \>56               \>       \> 1.4$  \times 10^{- 6 }$        \\
  50        \>   5$  \times 10^{- 9 }$           \> 56            \>         \>2.8$  \times 10^{- 7}$

  \end{tabbing}
\hrulefill \\

  \newpage
\section{Conclusions}

Of course particles with higher energy will be more penetrating than those of lower energy. This aspect of transport is not considered in the present paper. All that is considered is the product of the cross section with flux, which we called importance.
We have  shown that the small meson  production cross sections near threshold can be just as important as larger cross sections at  higher energy when transporting cosmic ray particles, because the shape of the cosmic ray spectrum enhances low energy cross sections, as used in the Boltzmann transport equation. 
In proton-proton reactions, this is true for pion production and especially true for kaon production. This leads us to speculate that  near threshold production of heavier hadrons  may also need to be  included in cosmic ray transport codes, 
because their thresholds are all above the pion threshold and the cosmic ray spectrum falls steadily as energy increases.

On the other hand this paper has shown that subthreshold pion production in nucleus-nucleus collisions is not important for cosmic ray transport because the energy of these reactions occurs below the peak of the nuclear cosmic ray spectra. However sub-threshold kaon production probably may need to be included in transport codes and again we speculate that sub-threshold production of heavier hadrons may need to  be included.

It is important to note that the present paper does {\em not prove} that the reactions considered {\em must} be included in all cosmic ray transport codes. All we have done is to suggest that certain reactions with small cross sections at low energy have similar importance factors compared to larger cross sections at higher energy, and therefore, under certain circumstances, these {\em might} need to be included in transport codes. The circumstances will depend on the incident spectrum, the particles one is transporting, the medium through which transport is taking place  and the particles of interest after transport through the medium.\\

\noindent Acknowledgements: 
JWN  and LWT were supported by
NASA grant NAG8-1901.  RBN was supported by NASA grant NNL05AA05H.

\end{document}